\documentclass{emulateapj}
\lefthead{PEPPER \& GAUDI} \righthead{DETECTING TRANSITING HOT EARTHS}
\def\eq#1{equation (\ref{#1})}

\def\days{\rm days}

\def\deg{^\circ}

\def\pt{{\cal P}_{tr}}
\def\pw{{\cal P}_{W}}
\def\psn{{\cal P}_{S/N}}

\def\ptot{{\cal P}_{tot}}

\def\ndet{N_{det}}
\def\sn{{\rm S/N}}
\def\snmin{(\sn)_{\rm min}}

\def\cd{{\cal D}}

\begin{document}

\title{Toward the Detection of Transiting Hot Earths and Hot Neptunes in Open Clusters}
\author 
{Joshua Pepper\altaffilmark{1} and B.\ Scott Gaudi\altaffilmark{2} }
\altaffiltext{1}{Ohio State University Department of Astronomy, 4055 McPherson Lab, 140 West 18th Ave., Columbus, OH 43210}
\altaffiltext{2}{Harvard-Smithsonian Center for Astrophysics, 60 Garden St., Cambridge, MA 02138}
\email{pepper@astronomy.ohio-state.edu,sgaudi@cfa.harvard.edu}

\begin{abstract}
Radial velocity searches for extrasolar planets have recently detected
several very low mass ($7-20M_\oplus$) planets in close orbits with
periods $\la 10~\days$.  
We consider the prospects for
detecting the analogs of these planets in Galactic open clusters via
transits.  We outline the requirements for constructing a transit
survey that would allow one to probe such ``Hot Earths'' and ``Hot
Neptunes.''  Specifically, we present a simple criterion for detection
that defines the minimum aperture required to detect planets of a
given radius in a cluster at a given distance.  Adopting photometric
precisions that have been demonstrated in
state-of-the-art variability surveys, we then predict the number of
planets one could potentially detect with ambitious transit surveys
toward several open clusters.  Dedicated surveys lasting more than 20
nights with Pan-STARRS toward the Hyades and Praesepe could detect a
handful of Hot Earths, if the majority of stars host such planets.
Similar surveys with larger aperture telescopes (e.g.\ CFHT, MMT),
toward M67, M35, M50, and M37 could detect Hot Neptunes, provided that
their frequency is $\ga 1\%$.  The majority of planets will be
detected around M dwarfs; detecting Hot Neptunes around such
primaries requires photometric precisions of $\sim 1\%$, whereas Hot
Earths require $\sim 0.1\%$.  We discuss potential hurdles in 
detecting and confirming small planets in ground-based surveys,
including correlated noise, 
false positives, and intrinsic stellar variability.

\end{abstract}
\keywords{open clusters and associations -- techniques: photometric -- surveys -- planetary systems}

\bigskip

\section{Introduction}

Over the past decade, the precisions of radial velocity (RV) searches
for extrasolar planets have steadily increased to the point where
several groups are currently achieving single-measurement Doppler
precisions of $\sim 1~{\rm m~s^{-1}}$ for quiet stars on a routine
basis \citep{mayor03,marcy05}.  As a result, RV searches have recently
detected several very low mass ($7-10~M_\oplus$) planets with periods $P
\simeq 2-10~\days$ \citep{santos04,mcarthur04,butler04,rivera05,lovis06}.
Detection of close-in planets with mass as low as $\sim M_\oplus$ 
may be possible using current technology \citep{narayan05}.

The origin and nature of these low-mass, short-period planets (``Hot
Earths'' and ``Hot Neptunes'') is not clear.  The more massive planets
could be ice giants similar to our own Neptune and Uranus that have
migrated to their current positions
\citep{il05}.  Rocky planets can form as agglomerations of
planetesimals that have been herded into short-period orbits via
sweeping resonances from migrating Jupiter-mass planets \citep{fogg05,zhou05}.
Jupiter-mass gas giants can be atrophied to Neptune mass or smaller
via photoevaporation \citep{baraffe05} or Roche lobe overflow caused
by unfettered inward migration \citep{trilling98} or
excessive internal heating \citep{gu03}.  

It is difficult to distinguish between these various scenarios with a
mass measurement alone.  However, constraints on the radii of these
planets may allow one to rule out some of the proposed hypotheses.
The $\sim 10\%$ of RV-detected companions that also happen to transit
their parent stars will allow for radius measurements, however
relatively large numbers of detections will be needed to detect many
transiting planets.

Transit searches have to date discovered seven planets.  Five of these
were discovered in deep field surveys of Galactic disk stars (see
\citealt{udalski04} and references therein), while two were discovered in
shallow, wide-angle surveys of nearby, bright stars
\citep{alonso04,mccullough06}.  None of these surveys were very
sensitive to planets with radii much smaller than Jupiter.  Transit
searches toward open clusters have a number of
advantages over surveys in the field \citep{janes96,vb05,pepper05}.
However such searches have not detected any planets, despite a large number of
completed and ongoing surveys (see \citealt{pepper05} and references
therein).  This is partly due to the small number of target stars.  However, Hot
Earths and Hot Neptunes may be more ubiquitous than Hot Jupiters, and
thus transit surveys toward open clusters with sensitivity to
smaller planets may meet with more success.

Here we consider the prospects for the detection of Hot Earths and Hot
Neptunes via transit surveys toward Galactic open clusters.  Our goal
is to show that detection of sub-Jovian sized planets may be possible
from ground-based facilities using current technology.  Previous transit 
searches have generally used simple, first-order calculations to make crude 
estimates for the number of expected transit detections, and were unable 
to predict the number of detections as a function of radius.  A comprehensive 
model for predicting transit detections with respect to all the relevant 
parameters is clearly required.  We have
previously developed just such a quantitative model of transit
searches toward stellar systems, which allows one to predict realistic
planet detection rates (\citealt{pepper05}, henceforth PG).  In this
paper, we apply this framework to first outline the criteria for
detection of planets with a given radius and period in a cluster of a
given distance (\S\ref{sec:detect}).   Adopting a set of reasonable, yet
optimistic assumptions, we then apply our results to
well-studied open clusters to predict the number of planets that
specific transit surveys would detect as a function of the radius of
the planets (\S\ref{sec:total}).  In section \S\ref{sec:real}, we
consider potential real-world difficulties in achieving our predicted
detection rates.  We summarize and
conclude in \S\ref{sec:conc}.

Our primary conclusion is that, by conducting ambitious, long-duration
surveys of Galactic open clusters using current or near-future
ground-based facilities, and employing state-of-the-art techniques in
relative photometry, it should be possible to detect transiting Hot
Neptunes and perhaps even Hot Earths.  The assumptions we adopt to
arrive at this conclusion are admittedly optimistic -- but not
unreasonable.  There are significant obstacles to this endeavor, namely 
reducing systematic errors, dealing with false positives, and observing 
stars with low intrinsic stellar variability.  We address these 
topics in \S\ref{sec:real}, and argue that they are potentially solvable.

In other words, the requirements for detecting such
small planets from the ground are certainly challenging, but do not
appear impossible.  Given the enormous potential payoff, we feel that
it is timely to consider the experiments we have proposed here in
greater detail, and to critically examine our assumptions.  Our goal
with this paper is to provide a stimulus for such explorations.

\section{Criteria for Detectability}\label{sec:detect}

In PG, we developed a model for transit surveys toward stellar systems
(e.g.\ open clusters) that allowed us to predict the number of planets
$\ndet$ that a particular survey would detect as a function of the
parameters of the system, the observational setup, site properties,
and planet properties.  We refer the reader to that paper for an
in-depth discussion of the model, its assumptions, and ingredients.
The basic ingredient in the estimation of
$\ndet$ is $\ptot(M,P,r)$, defined as the probability that a planet of
radius $r$ and orbital period $P$ will be detected around a star of
mass $M$.

The detection probability $\ptot$ can be separated into
three factors (\citealt{gaudi00}, PG),
\begin{equation}\label{eqn:ptot}
\ptot(M,P,r) = \pt(M,P) \psn(M,P,r) \pw(P).
\end{equation}
Here $\pt$ is the probability that the planet transits its parent
star, $\psn$ is the probability that an observed transit will yield a
signal-to-noise ratio (S/N) that is higher than some threshold value,
and $\pw$ is the window function which describes the probability that
at least two transits will occur during the survey and so enable
an estimate of the period.  The transit probability is $\pt = R/a$, where
$R$ is the stellar radius, and $a$ is the planet semimajor axis.

The function $\pw(P)$ is just the probability that a planet with a
given $P$ will exhibit at least two transits during the observations.
This function depends on the total number of nights observed $N_n$, the
duration of each night $t_{night}$, and $P$.  Assuming perfect weather and $t_{night}=8~{\rm
hr}$, we find that a campaign must last at least $N_n \sim 15$
nights in order that $\pw>80\%$ for $P=1-4~\days$, assuming a
uniform distribution in $\log P$.  Accounting for weather, we advocate
$N_n \sim 20$ nights as a minimum duration for transit campaigns.  

In PG we demonstrated that, for typical parameters, $\psn$ is maximized for 
observations in the $I$-band.  Furthermore, at fixed $(r,P)$
the $\sn$ in the $I$-band is weakly dependent on the mass of the primary for sources with flux above
the sky background, whereas the $\sn$ falls sharply for sources below sky
(see \S 3.2 of PG).  As a result, if it is possible to detect
planets with a given $(r,P)$ around
stars with flux equal to the sky background, then it is possible to detect
such planets around all brighter stars in the cluster. 
Therefore, we can construct a simple
``detectability index'' for deciding if a particular experiment
is capable of detecting planets around
stars in a particular cluster. This index is simply the criterion
that a planet with $(r,P)$ would give rise to a transit with $\sn$
greater than some threshold value $\snmin$ around a star with flux equal
to sky.  Assuming power-law forms for the stellar mass-radius and mass-luminosity 
relations, the detectability 
index is (PG)
\begin{equation}\label{eqn:dindex}
\cd= 2C_1C_2^{(3\alpha-\beta_\lambda+1/3)/\beta_\lambda},
\end{equation}
where $C_1$ and $C_2$ are given by,
\begin{equation}\label{eqn:c1}
C_1 = 
(1024\pi)^{1/3}[\snmin]^2
\left(1+\frac{t_{read}}{t_{exp}}\right)
\left(\frac{r}{R_\odot}\right)^{-4} 
\end{equation}
$$
\times \left(\frac{d}{D}\right)^{2} 
\left(\frac{GM_\odot}{P R_\odot^3L_{\lambda,\odot}^3}\right)^{1/3}
10^{0.4A_\lambda},
$$
\begin{equation}\label{eqn:c2}
C_2 =
\frac{4\pi d^2 S_{sky,\lambda}\Omega}{L_{\lambda,\odot}10^{-0.4A_\lambda}}.
\end{equation}
Here $t_{read}$ is the detector readout time, $t_{exp}$
is the exposure time, $d$ is the distance to the cluster, $D$ is the
telescope aperture, $L_{\lambda,\odot}$ is the photon luminosity of
the sun, $A_{\lambda}$ is the extinction toward the cluster,
$S_{sky,\lambda}$ is the photon surface brightness of the sky, $\snmin$ is the minimum $\sn$ required for detection, and
$\Omega=(\pi/\ln 4)\theta_{see}^2$ is effective area a PSF with FWHM
$\theta_{see}$. The subscript $\lambda$ denotes bandpass-specific
quantities. 
The variables $\alpha$ and $\beta_{\lambda}$ are the power-law indices
for the mass-radius and mass-luminosity relationships.

\begin{figure}
\epsscale{1.0}
\plotone{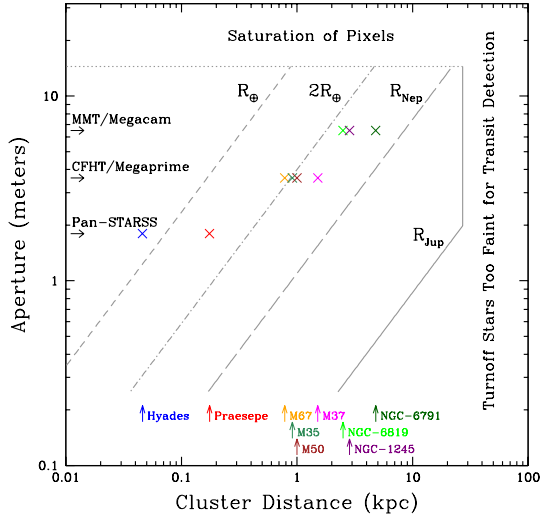}
\caption{Aperture sizes necessary to detect planets in clusters of varying
distances.  The curves show detectability regions in the $I$-band for four planet sizes.  
Apertures larger than the respective lines are required for
detectability.  Above the dotted line the sky will saturate the detector
in $t_{exp}=45~{\rm s}$. The crosses indicate the aperture sizes
for each cluster used to generate the curves in Figure \ref{fig:ndet}.}
\label{fig:1}
\end{figure}

When $\cd \leq 1$, a survey can successfully detect planets with 
($r,P$) around all stars with flux above sky.
Figure \ref{fig:1} shows the value of $D$ for which $\cd=1$ as a
function of $d$ for planets with radius equal to
Earth, twice Earth, Neptune, and Jupiter ($r=R_\oplus, 2R_\oplus, R_{Nep}, R_{Jup}$) 
and $P=2~\days$.  Apertures
above the curves yield robust detections of planets with the given $r$, while below the curves,
the number of detections falls rapidly.  We
have assumed $I$-band observations, $t_{read} = 15~{\rm s}$, $t_{exp} = 45~{\rm s}$,
$\snmin = (30)^{1/2}, A_I=0.2$, $\theta_{see}=1''$, and
$S_{sky,I}=19~{\rm mag/arcsec^2}$.  We can use Figure
\ref{fig:1} to determine how large a telescope is required to
detect a planet of a given radius in a target cluster of a given
distance.  The cutoff at large distances is due to the fact that at
such distances the turnoff stars (assuming a cluster age of 1
Gyr) have flux below the sky background.  The 
cutoff at large $D$ is due to the fact that, for sufficiently large apertures, the
sky itself will saturate the pixels in $t_{exp}$, assuming pixels of
angular size $\theta_{pix}=0.2''$ and full well depth of $N_{FW}=10^5$
electrons.  Figure \ref{fig:1} also shows the distances to several
potential target clusters.

One factor that deserves comment is the exposure time.  From
\eq{eqn:c1}, it is clear that one is driven to longer exposure times
to avoid wasting too much observing time on readout, and that as long
as $t_{exp}$ is significantly longer than $t_{read}$, the
detectability is roughly independent of $t_{exp}$.  On the other hand,
one is driven to short exposure times to avoid saturating on the
brighter cluster stars.  Therefore, there is an optimal exposure time
which will depend on the cluster distance, age, and telescope
aperture.  For simplicity, we will fix $t_{exp}=45~{\rm s}$ unless otherwise
indicated, but we
note that this value is not necessarily optimal for all setups.

\section{Predictions for the Number of Detections by Cluster}\label{sec:total}

We can calculate the total number of expected transit detections
$\ndet$ by convolving $\ptot$ over the mass function of the cluster and planetary frequency
distribution as a function of $(r,P)$ (see Equation 1 of PG).  We
adopt a power-law mass function of the form ${dn}/{dM}\propto
M^\gamma$ for $0.3M_\odot \le M \le M_{to}$, where
$M_{to}$ is the turnoff mass, as determined from the age of the
cluster.  We do not consider stars with $M\le 0.3M_\odot$,
because the mass function slopes below this mass are generally poorly
known. We assume that every star has a planet of a given radius,
distributed uniformly in $\log P$ between $1-4~\days$.
Since we assume that every star has a planet with the 
given radius, the total detection numbers must be multiplied by the actual 
fraction $f$ of stars with such planets.

We now estimate $\ndet$ for specific realizations
of transit surveys toward well-studied Galactic open
clusters.\footnote{Here we focus on clusters in the northern
hemisphere, mainly because these are the most well-studied clusters.  
See \citet{vb05} for a list of potential southern
hemisphere open cluster targets.}  For all clusters, we assume
$I$-band observations, $N_n=20~\days$ (no bad weather), $t_{night}=8~{\rm hr}$,
$t_{read} = 15{\rm s}$, $\snmin = (30)^{1/2}$,
$\theta_{see}=1''$, $S_{sky,I}=19~{\rm mag/arcsec^2}$,
$\theta_{pix}=0.2''$, and $N_{FW}=10^5$.  For all the 
clusters we assume $t_{exp} = 45{\rm s}$, except for the Hyades,
where we assume $15{\rm s}$.  We also assume a minimum
photometric error of $\sigma_{sys}=0.1\%$, 
to account for systematics.  Millimagnitude
precisions have been demonstrated with state-of-the-art 
reduction methods using image-subtraction photometry (see, e.g., \citealt{hartman05}).
Our choice is therefore realistic, although perhaps optimistic. 
In \S\ref{sec:real}, we discuss the appropriateness of this choice in detail, 
and consider the effects of less optimistic
choices on our conclusions.

\begin{figure}
\epsscale{1.0}
\plotone{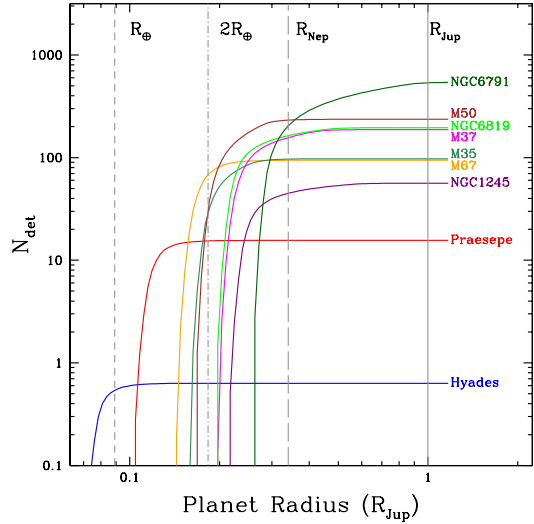}
\caption{
The number of planets detected, $\ndet$, versus planet radius for
various Galactic open clusters, assuming every cluster member has a
planet of the given radius. In order to arrive at the actual
number of detections, these curves must be multiplied by the
fraction of stars with such planets. We have assumed cluster parameters given
in Table \ref{tab:1}, experimental parameters given in the text, and
that planets are distributed uniformly in $\log P$ between 1
and 4 days.  The aperture sizes used for each 
cluster are shown in Figure \ref{fig:1}.}
\label{fig:ndet}
\end{figure}

For each cluster, we must also specify $d$, $\gamma$, $A_I$, 
age, and the total number of stars $N_*$ in the range $0.3M_\odot \le M \le M_{to}$.  We 
searched the literature for well-studied open
clusters for which these values have been estimated.  We
assume $R_V=3.1$ and $A_I/A_V = 0.482$ \citep{bm98}.  We chose to
include following clusters: the Hyades, Praesepe, M67, M35, M50, M37,
NGC 6819, NGC 1245, and NGC 6791.  The relevant parameters from these
clusters and references are listed in Table \ref{tab:1}.  They span
distances from $d=46~{\rm pc}$ to $4.8~{\rm kpc}$, and richnesses
from $N_* \simeq 200$ to $5600$.

\begin{deluxetable*}{cccccc|cccc}
\tablecaption{Cluster Properties and Transit Detection Rates}
\tablewidth{0pt}
\tabletypesize{\scriptsize}
\tablehead{
  \colhead{Cluster} &
  \colhead{$d$} &
  \colhead{Age} &
  \colhead{$N_*$}\tablenotemark{a}&
  \colhead{$A_I$} &
  \colhead{$\gamma$}&
  \multicolumn{4}{c}{$N_{det}/f$\tablenotemark{b}}\\
  \colhead{} &
  \colhead{(pc)} &
  \colhead{(Myr)} &
  \colhead{} &
  \colhead{(mag)} &
  \colhead{} &
  \colhead{$R_\oplus$} &
  \colhead{$2R_\oplus$} &
  \colhead{$R_{Nep}$} &
  \colhead{$R_{Jup}$}}
\startdata
Hyades\tablenotemark{1,2,4} 	&   46 	&  625 	&  180	& 0.002 & -0.7 & 0.54 & 0.63 & 0.63 & 0.63 \\
Praesepe\tablenotemark{3,4} 	&  175 	&  800 	&  570 	& 0.03 & -1.63 & -- & 15.4 & 15.6 & 15.6 \\ 
NGC 2682 (M67)\tablenotemark{5}	&  783	& 4000	& 2150	& 0.07 & -0.51 & -- & 61.7 & 94.1 & 94.2 \\
NGC 2168 (M35)\tablenotemark{6} &  912 	&  180 	& 1500 	& 0.30 & -2.29 & -- & 20.6 & 96.6 & 97.4 \\ 
NGC 2323 (M50)\tablenotemark{6} & 1000 	&  130 	& 3200 	& 0.33 & -2.94 & -- & 16.0 & 231.7 & 236.3 \\ 
NGC 2099 (M37)\tablenotemark{7} & 1513 	&  580 	& 2600 	& 0.35 & -1.60 & -- & -- & 156.5 & 187.5 \\
NGC 6819\tablenotemark{8} 	& 2500 	& 2500 	& 2900 	& 0.15 & -0.85 & -- & -- & 163.8 & 195.5 \\ 
NGC 1245\tablenotemark{9}	& 2850 	& 1000 	& 960 	& 0.33 & -0.5 & -- & -- & 44.9 & 56.3	\\ 
NGC 6791\tablenotemark{10,11} 	& 4800 	& 8000 	& 5600 	& 0.15 & -1.30 & -- & -- & 204.1 & 535.6 \\ 
%
\enddata
\tablenotetext{a}{Number of stars between the turn-off and $0.3M_\odot$, 
as derived from referenced sources and recalibrated for the 
specified mass range.}
\tablenotetext{b}{Total number of detected planets, divided
by the fraction $f$ of stars with planets with the indicated radius and $P=1-4~\days$.}
\tablerefs{ 
(1) \citet{perry98}; 
(2) \citet{reid99}; 
(3) \citet{adams02}; 
(4) \citet{tay02};
(5) \citet{fan96};
(6) \citet{kal03};
(7) \citet{kal01b};
(8) \citet{kal01a};
(9) \citet{burke04};
(10) \citet{chab99};
(11) \citet{kaz92}}
\label{tab:1}
\end{deluxetable*}

As is clear from Figure \ref{fig:1}, the choice of aperture 
can have large effect on the smallest detectable planets.  The
largest available apertures are not necessarily always indicated,
both because of the trade-off between saturation and efficiency, as
discussed in \S\ref{sec:detect}, and because closer clusters, which
have diameters of many degrees on the sky, require large
fields-of-view, which are generally easier to construct on smaller $D$
telescopes.  We consider three different telescope/detector
combinations:
Pan-STARRS\footnote{http://pan-starrs.ifa.hawaii.edu/public/index.html;
\citet{kaiser02}} ($D = 1.8{\rm m}$, FOV$={3\deg}\times{3\deg}$),
CFHT/Megaprime\footnote{http://www.cfht.hawaii.edu/Instruments/Imaging/Megacam;
\citet{boulade98}}
($D=3.6{\rm m}$, FOV$=1\deg \times 1\deg$), and
MMT/Megacam\footnote{http://cfa-www.harvard.edu/cfa/oir/MMT/MMTI/megacam.html;
\citet{mcleod00}} ($D=6.5{\rm m}$, FOV$=24' \times 24'$).  We adopt
$D=1.8{\rm m}$ for the Hyades and Praesepe, $D=3.6{\rm m}$ for M67,
M35, M50, and M37, and $D=6.5{\rm m}$ for NGC 6819, NGC 1245, and NGC
6791.  We assume that the clusters fit in the FOV in all cases,
although for the Hyades and Praesepe this implies that all four
Pan-STARRS telescopes will need to monitor the clusters simultaneously due to
their large angular size ($\ga 5\deg$).

The predictions for $\ndet(r)$ for the clusters are shown in Figure
\ref{fig:ndet} and tabulated in Table \ref{tab:1} for $r=R_\oplus, 2R_\oplus,
R_{Nep}$ and $R_{Jup}$.   Note that the placement of 
cluster/aperture combinations in Figure \ref{fig:1} demarcate where detections 
at that radii level off; planets can be detected below 
their radius curves in Figure \ref{fig:1}, but in rapidly decreasing numbers.

\section{Sensitivity to Real-World Effects}\label{sec:real}

The numbers we present in the previous section are based on
assumptions that, while reasonable, may be considered optimistic.  We
therefore take another look at three assumptions: the window function
and assumptions about weather, the number of transits required to
confirm a detection, and the minimum photometric error.  We also
discuss the effects of correlated noise and intrinsic variability on the ability to
reliably detect small transiting planets, and 
discuss the problem of false-positive detections.

\subsection{Window Function}
For the purposes of obtaining a generic analysis, we assumed in \S\ref{sec:total} that 
each cluster is observed for 20 consecutive cloudless nights.  That is obviously an 
idealized situation.  Here we test the sensitivity of the number of expected
detections to the amount of time lost to weather.  Poor weather affects
the number of detections through the window function $P_W$, which gives
the probability that two or more transits will occur during
observations.  The window function also depends on the number of transits
required, the length of the nights, and the observing strategy.  We consider
the effects of requiring three transits in the following section, but we do
not attempt to quantify the effects of the observing strategy
or the length of the nights.  See \citet{vb05} for a thorough exploration
of the effects of these factors on the window function.

In order to test how vulnerable our results are to bad weather,
we calculate additional window functions, which combine a number of nights that are 
completely lost to weather with a number of nights that are partially lost.  We utilized 
two scenarios, one representing moderate weather loss (3 nights lost, 3 nights 
partially lost), and one representing severe weather loss (7 nights lost, 3 nights 
partially lost).  For nights with partial 
loss to weather, a random block of time lasting between zero and five hours is lost at a 
random point in the night.  We also model two types of weather behavior, where the 
lost nights are clustered into a block, and where they are randomly scattered through 
the length of the survey.  In both cases the partially lost nights are scattered randomly.

\begin{figure}
\epsscale{1.0}
\plotone{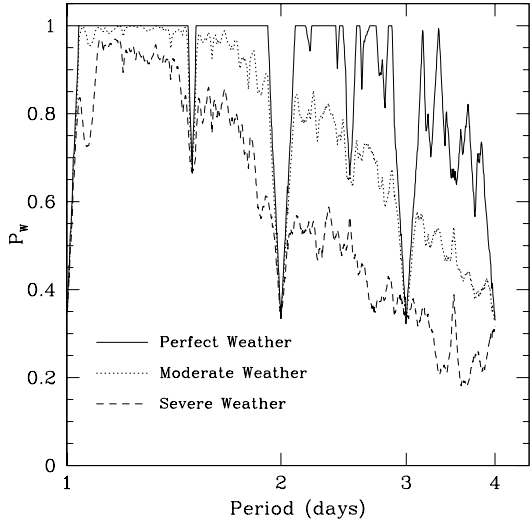}
\caption{Different weather patterns affect the rate of transit detections at various periods.  Here 
we show the effects of moderate weather (dotted line) and severe weather (dashed line) as compared 
with perfect weather (solid line).  The weather-affected lines are each the average 
of 100 sample window functions using the definitions of moderate and severe weather as described 
in the text.}
\label{fig:wins}
\end{figure}

We found that clustering the bad weather in time did not significantly reduce the overall 
number of planets discovered.  However, we decided to keep the lost nights clustered, since 
that pattern most closely approximates the conditions of real loss to 
weather.  Figure \ref{fig:wins} shows three sample window functions.  The solid line is the 
window function for perfect weather.  The dotted line is the window function for an average 
of 100 cases of moderate weather loss.  The dashed line is the window function for an 
average of 100 cases of severe weather loss.

The overall effect of a moderate loss of time due to weather 
is relatively minor $11\%$ reduction in the number of detections.  
Severe weather reduced the detections by $36\%$.  These losses naturally  
occur preferentially for planets with longer periods.  

\subsection{Number of Required Transits}

In \S\ref{sec:total} we required that two transits be observed for a
planet detection.  The requirement of multiple transits aids in the
elimination of false positives due to systematics, and enables an
estimate of the period of the transiting companions.  However, two
transits may not be sufficient to reliably constrain the period of the
planet, and some authors have argued that the detection of three transits
is absolutely essential to exclude false positives and so claim a robust
detection.  We therefore also consider the effect of requiring that at least three
transits be observed for detection.

In our model, requiring additional transits for a detection affects
the detection rates by changing the window function, similar to
including bad weather.  We find that, for perfect weather, requiring
three transits reduces the detection rate by $16\%$ as compared to
requiring two transits.

\subsection{Photometric Error}

The ability to achieve low levels of photometric errors over the
course of an entire observing run is crucial to the success of transit
surveys.  Consistently low photometric errors are required to detect
small transits.  The fact that some surveys have not been able to
achieve the predicted error levels has slowed the progress of transit
detection projects.

It is beyond the scope of this paper to present a detailed discussion
of what is required to achieve the millimagnitude photometric precisions we have
adopted here, and under what circumstances one can expect such
precisions to realistically be achieved.  However, it is worthwhile to
briefly consider the current state-of-the-art in relative
photometry.  \citet{hartman05} have demonstrated it is possible to
achieve very high photometric precisions on a large number of stars
using a large-format detector on a large telescope, combined with
image-subtraction photometry.  In particular, their photometry exhibit
a systematic floor of one millimagnitude, and they show that this
precision is stable over time scales of several days.  Although it is
unclear whether these results can be achieved with other
observational setups, we note that there is nothing particularly
special in the setup used by \citet{hartman05} that allowed them to achieve these
levels of precision (except for perhaps the
small pixel size of the detector).  The primary improvement over previous work comes
from more careful data reduction treatments.  Thus there is no reason
to believe that the results of \citet{hartman05} cannot be replicated
in other experiments.

Nevertheless, it is interesting to ask
how sensitive our predictions are to the assumed minimum photometric
error.  We note that, in our simulations, the majority of planets are
detected around low-mass M-dwarfs with
$R\sim 0.3R_\odot$, for which the transit depth is $\sim 0.1\% (r/R_\oplus)^2$.
Thus we can expect that, in order to detect Hot Earths, we require 
photometric precisions of $\sim 0.1\%$.  However, the transit depth for a Neptune-sized
planet orbiting an M dwarf is closer to $\sim 1.5\%$.  Therefore,
we can expect that photometric precisions of $\sim 1\%$ should still
be sufficient to detect Hot Neptunes.   We confirm these expectations
by recalculating the detection rates for minimum errors
between $\sigma_{sys}=0.1\%$ and $1\%$.  We indeed find that the detection Hot Earths
requires photometric precisions of $\sim 0.1\%$. 
At $\sigma_{sys} = 1\%$, we find that Neptune-sized transits can still
be detected in the five nearest open clusters considered in this
paper.  Such precisions should be routinely achievable using most
setups.
 
\subsection{Correlated Noise and Systematics}

An implicit assumption in our calculations is that the photometric
errors are not temporally correlated.  As discussed by \citet{pont05},
photometric errors that are
correlated on the same time scale as the typical duration of a planetary
transit (a few hours) can have a dramatic effect on the
ability to reliably detect weak photometric signals.  Indeed, analysis of the photometry from
the OGLE collaboration which led to the detection of the first planets
via transits indicates that such correlated errors are present and can
be quite important \citep{gould06}.  These correlated errors not only
increase the minimum required signal-to-noise ratio for detection, but
affect how this minimum signal-to-noise ratio depends on period and
source star brightness \citep{pont05}.  The existence of these errors
has led \citet{pont05} to argue that Neptune-sized planets cannot be
detected from the ground.  However, various algorithms have been developed 
that may allow one to control or remove these correlated errors 
entirely \citep{tamuz05,kovacs05,burke05}.
Indeed, analysis of the photometric data presented in \citet{hartman05}, which applied one of these algorithms, 
shows no evidence for correlated errors at a significant level
(J.\ Hartman, private communication).  Nevertheless, it is clear that this is an important
issue, and caution is warranted, especially in the search for lower-amplitude transit signals.

\subsection{False Positives and Follow-up}

The complexity of confirming transit detections has bedeviled all 
transit searches.  There are a number of phenomena that can mimic 
photometric transit signals.  All of these false positives can be eliminated by sufficient 
precision RV follow up.  However for the majority of planets detected by the surveys we 
consider here, the host stars will be too faint for precision RV follow up.  Fortunately, as 
outlined by \citet{brown03}, most false positives are more important for transit 
depths of 1\% or more, and become less common for smaller transit depths.  Furthermore, the one 
source of false positives that is still common at small transit depths, namely blending of main 
sequence binaries with a foreground or background star, is more important for wide 
angle surveys with large point spread functions where blending is common.  For 
the surveys we consider here, blending is much less likely.

Regardless, we are entering the era of large, space-based 
surveys (e.g. COROT\footnote{http://smsc.cnes.fr/COROT/} and Kepler\footnote{http://kepler.nasa.gov/}) which 
aim to detect planets for which RV follow-up is difficult or impossible.  The ground-based 
surveys which we envision here will prove important for identifying possible sources 
of false positives which will be encountered in such second generation surveys.  Thus 
the potential difficulties faced by these surveys are not unique to this class 
of planets searches.

\subsection{Intrinsic Variability}

Intrinsic stellar variability may overwhelm any signals from the low 
amplitude transits we are searching for here.  A number of 
transit searches have characterized stellar variability down to M 
dwarfs (e.g.\ \citealt{hartman05}).  For stars older than $\la 200~{\rm Myr}$ significant 
intrinsic variability has not been observed at the $\ga 1\%$ level.  For systems younger 
than 200 Myr, such variability has been observed, which would make these surveys 
difficult or impossible.  At the mmag level at which Neptunes could be detected, most solar type
stars show no evidence for variability on the time scales and with the
duty cycles characteristic of planetary transits.
Little is known of the intrinsic 
variability of less massive stars, such as M stars, at these levels.  If such variability exists, then 
all surveys that aim to detect sub-Jovian planets with transits will encounter difficulties.

\section{Discussion and Conclusions}\label{sec:conc}

We have used a model to show that dedicated transit surveys in clusters 
have the potential to detect Neptune- and even Earth-sized extrasolar 
planets.  Surveys with Pan-STARRS
toward Hyades or Praesepe would be able to detect planets with $P\la
4~\days$ and radii as small as the Earth.  If a fraction $f$ of
stars host Hot Earths with $r=2R_\oplus$ and $P=1-4~\days$, than these
surveys would detect $\sim 0.5f$ toward Hyades and $\sim 15f$ toward
Praesepe.  Surveys toward more distant clusters with larger aperture telescopes
such as the CFHT or MMT would be sensitive to Hot Neptunes, even if
they are relatively rare with $f \sim 1\%$.  For example, a 20-night
survey with CFHT toward M37 would detect $\sim 150f$ Hot Neptunes (and
$\sim 190f$ Hot Jupiters).  

There are a number of effects that could reduce our predicted 
numbers.  The detection numbers from Table \ref{tab:1} would be lower by a factor of 
about $27\%$ if we assume moderate weather patterns and require three transits 
to confirm a detection.  We have 
also assumed a fairly low threshold for detection
($\sim 5.5\sigma$), and no correction for binaries.  We 
have also used a boxcar-shaped transit curve model, thus ignoring
the ingress/egress durations and limb-darkening of the stars.  
Estimates from \citet{burke05} suggest 
that such real-world effects would reduce detection rates by factors of $\sim 1.5-2$. 
However, we have made other assumptions which are somewhat
conservative.  For example, we have assumed detection thresholds based
on the $\sn$ of one transit -- it is possible to improve the $\sn$ for
multiple transits by folding the observed light curve about the
appropriate phase (see Appendix A of PG).  It is also possible to improve the
detection rates by simply increasing the duration of the survey beyond
20 nights.

Most of the planets detected in these surveys would be found around low-mass M-dwarfs with
$R\sim 0.3R_\odot$, for which the transit depth is $\sim 0.1\% (r/R_\oplus)^2$; this
is what allows the detection of such small planets from the ground,
despite the inevitable systematic errors in the relative photometry,
which we assumed to be $\sim 0.1\%$.  We have argued that 
this value is reasonable; however, if the minimum photometric error were larger 
than $0.1\%$, the surveys would miss the lower-radius planets. 
We note that the detection of Hot Neptunes
does not rely critically on achieving such a low systematic error. 
The transit depth for a Hot Neptune orbiting an M-dwarf is $\sim 1.5\%$. 
Thus we find that even if we assume a very conservative 
systematic error of $0.5\%$, the detection rates for Hot Neptunes are
mostly unaffected.

We have discussed how two potential hurdles -- systematic noise and false 
positives -- should be manageable for these surveys.  The intrinsic 
variability of M dwarfs is a potential problem that might derail attempts 
to detect sub-Jovian planetary transits from the ground.  However, if that 
were to become a significant problem for the surveys we describe here, it 
would be just as severe a problem for more ambitious, space-based 
surveys, and as such should be explored sooner rather than later.

Thus we conclude it may be possible -- from the ground
and with current technology -- to place interesting constraints on the
frequency of Hot Earths and Hot Neptunes in Galactic open clusters.
This will in turn constrain the properties of the low-mass planets
recently detected in RV surveys, as well as theories of planet formation and
migration in general, and inform future searches for extrasolar planets.

\acknowledgments
We would like to thank Chris Burke, Marc Pinsonneault and Josh Winn for helpful
discussions.  This work was supported by a Menzel Fellowship from the
Harvard College Observatory, and also by the National Aeronautics and
Space Administration under Grant No. NNG04GO70G issued through the
Origins of Solar Systems program.

\end{document}